\documentclass[aps, prl, reprint, longbibliography]{revtex4-1}
\usepackage{graphicx,amsmath}
\usepackage{subfig}
\DeclareGraphicsExtensions{.pdf,.eps,.png,.jpg,.mps}
\graphicspath{ {Figures/} }

\begin{document}

\title{Coincidences with the Artificial Axon}
\author{Hector G. Vasquez}
\author{Giovanni Zocchi}
\email{zocchi@physics.ucla.edu}
\affiliation{Department of Physics and Astronomy, University of California - Los Angeles}

\begin{abstract}
\noindent The artificial axon is an excitable node built with the basic biomolecular components and supporting 
action potentials. Here we demonstrate coincidence firing (the AND operation) and other basic 
electrophysiology features such as increasing firing rates for increasing input currents. 
We construct the basic unit for a network by connecting two such excitable nodes through an electronic synapse,  
producing pre/post synaptic behavior in which one axon induces firing in another.  
We show that the system is well described by the Hodgkin-Huxley model 
of nerve excitability, and conclude with a brief outlook for realizing large networks of such low voltage ``ionics''. 
 
\end{abstract}

\maketitle

\noindent {\bf Introduction.} 
Electrical signal conditioning under water is the hallmark of nerve cells. Action potentials 
- electrical spikes in the neuron - result from two coupled nonlinear relaxation phenomena, involving 
``molecular transistors'' (voltage gated ion channels in the cell membrane) and supported by the 
non-equilibrium state created by ionic gradients maintained across the cell membrane. The fundamental 
mechanism of action potentials was elucidated by Hodgkin and Huxley in the 1950s, and the knowledge 
constructed since then through electrophysiology experiments on live neurons is monumental 
\cite{Koch, Hille}. 
\begin{figure}
\centering
\includegraphics[width=3in]{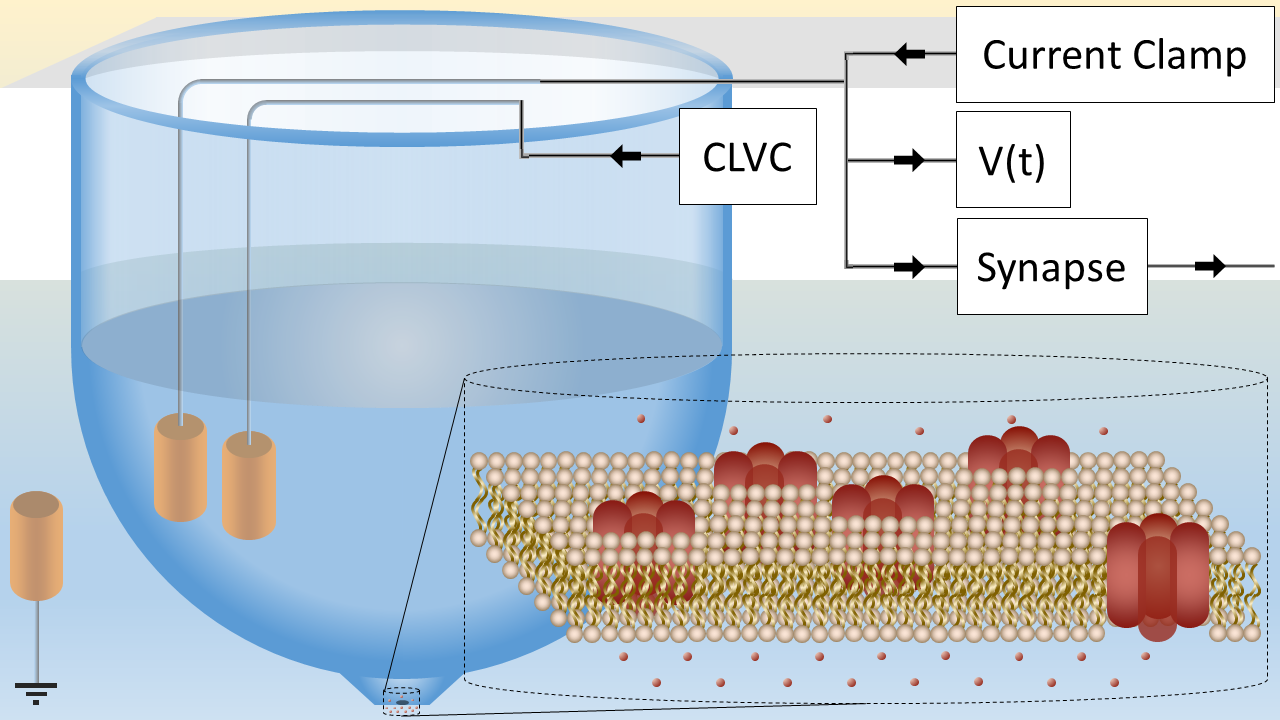}
\captionsetup{justification=raggedright, singlelinecheck=false}
\caption{Sketch of the artificial axon. A lipid bilayer painted on a $\sim100\mu m$ hole in a plastic tube separates two fluid compartments where different KCl concentrations are maintained. The membrane is studded with $10^2 - 10^3$ voltage-gated potassium channels (KvAP). The current limited voltage clamp (CLVC) holds the system off the equilibrium Nernst potential, fulfilling the role of the sodium ions and channels in the real axon. The electrode for measuring membrane voltage also serves as the output electrode for a current clamp and input for the electronic synapse, which is a current clamp connecting to the next axon.}
\label{fig:setup}
\end{figure}
And yet, action potentials were never taken out of the live cell into an in vitro setting. This most 
interesting of dynamical systems has remained confined to the living world. We recently produced 
a similar excitable medium using the biological components in an artificial setting \cite{Amila2016}. 
The difficulty of constructiong a functional membrane with two different, correctly oriented ion channel 
species was circumvented by the use of what we call the current limited voltage clamp (CLVC) and only 
one channel species. The system exhibits threshold behavior and signal amplification, the two hallmarks 
of action potentials. Here we use the CLVC to more specifically render the effect of a second ionic species 
and channel type, moving our artificial electrophysiology platform closer to a bona fide artificial axon. 
We inject currents using a current clamp in the traditional electrophysiology way and demonstrate 
multiple firing. We demonstrate coincidence detection: the AND operation, fundamental 
for biological neural networks \cite{Elisha2008}. Finally, we connect two axons with an electronic 
``synapse'' and demonstrate signal propagation and conditioning. We show that the dynamics of the 
system is captured by the classic Hodgkin - Huxley (HH) model. We conclude 
with the outlook for an extended network of artificial axons. \\ 
Reconstituted lipid membranes with and without channels have a history dating from the 1960s 
\cite{Mueller1962}; one important use today is to study ion channel dynamics \cite{Schmidt2009, Bassereau2011, Amila2013, MacKinnon2003}. The alternative method is patch clamping of real cells 
\cite{Sakmann1984}, which has also been used to study the dynamical system of a collection of channels 
\cite{Erez1, Erez2}. \\ 

\noindent {\bf Results.} In the real axon, two steady state, opposing gradients of $Na^+$ and $K^+$ ions, maintained 
across the cell membrane by ATP driven molecular pumps, and small but essential leak conductances 
for both ions, result in a steady state potential difference across the membrane (the ``resting potential'' 
$V_r$) which is neither the equilibrium (Nernst) potential for the sodium ions 
($V_N^{Na} \sim + 60 \, mV$) 
nor the Nernst potential for the potassium ions ($V_N^{K} \sim - 100 \, mV$). Action potentials are 
generated when synapses effectively inject a small current into the neuron, raising the intracellular 
potential $V$ (we refer potentials to the grounded ``outside'' of the cell). The voltage sensitive sodium 
channels open, forcing $V \rightarrow V_N^{Na}$. At these positive voltages the potassium channels 
open, while the sodium channels close (inactivate), pushing $V \rightarrow V_N^K$. Finally, the 
potassium channels close again, and $V \rightarrow V_r$. The width of the spike (a few ms for 
mammalian neurons \cite{Koch}) is governed by the interplay of several characteristic time scales; 
the amplitude is fixed by the ionic gradients (essentially $V_N^{Na} - V_N^K \sim 100 \, mV$). \\
\noindent The artificial axon \cite{Amila2016} is a similarly excitable supported phospholipid bilayer patch 
of size $\sim 100 \, \mu m$, separating two fluid compartments (Fig. 1). This artificial membrane is 
studded with $\sim 100$ oriented voltage gated potassium channels (KvAP), and a concentration 
difference $[K^+]_{in} \approx 30 \, mM$, $[K^+]_{out} \approx 150 \, mM$ is maintained between 
the inside and outside compartments, corresponding to an equilibrium (Nernst) potential: 
\begin{equation}
V_N \, = \, \frac{T}{|e|} ln \frac{[K^+]_{out}}{[K^+]_{in}} \, \approx \, + \, 40 \, mV
\label{eq:Nernst}
\end{equation}
We keep the system out of equilibrium, at the ``resting potential'' $V_r \approx - 110 \, mV$ where 
the KvAP channels are closed, using a voltage clamp (Fig. 1), which thus has the role of the second 
ionic gradient in a real axon. However, no dynamics can be generated if the voltage is really clamped, 
so we limit the current that can be sourced by the clamp using a resistance $R_c$ in series with 
the clamp electrode. $R_c$ is chosen so that the maximum current sourced by this ``current limited 
voltage clamp'' (CLVC) is larger than the leak current with channels closed, but smaller than the 
channel current with channels open. Then the system can still ``fire''. More details on the electronics 
and construction of the excitable membrane are found in \cite{Amila2016} and will be given in 
a longer publication in preparation. \\
Fig. 2 shows the response of the system to a sub-threshold stimulus (a) and a stimulus above 
threshold (b). The blue curve is the actual voltage $V(t)$ of the axon, the yellow curve the current 
$I_{CLVC}(t)$ through the clamp, and the dotted curve shows the clamp protocol $V_c (t)$ , which 
is a square pulse to $- 25 \, mV$. In (a) the effect is simply the charging and discharging of the 
membrane capacitance, with an $R C$ timescale given by the membrane capacitance $C \approx 192\,pF$ 
and the CLVC resistance $R_c = 100 \,M\Omega$. In (b) the square pulse stimulus is longer 
($100 \, ms$ instead of $50 \, ms$), and the response is completely different: the system ``fires''. 
Notice that after the clamp is pulled back down to $V_c = - 125 \, mV$ (at $t \approx 100 \, ms$) 
the axon voltage $V(t)$ continues to climb, because channels are open and the channel current 
overwhelms the clamp current. $V(t)$ does not quite reach the Nernst potential 
$V_N \approx + 40 \, mV$ because of this competition. The subsequent decrease of $V(t)$ is due 
to channel inactivation. These traces are essentially deterministic and can be repeated many times. \\ 
Two sub-threshold pulses in close proximity will cause the system to fire, implementing an AND operation 
or coincidence detector. An example is shown in Fig. 3 where each $86 \, ms$ long pulse alone 
is sub-threshold, but two such pulses $80 \, ms$ apart cause firing. \\ 
\begin{figure}
\centering
\includegraphics[width=3in]{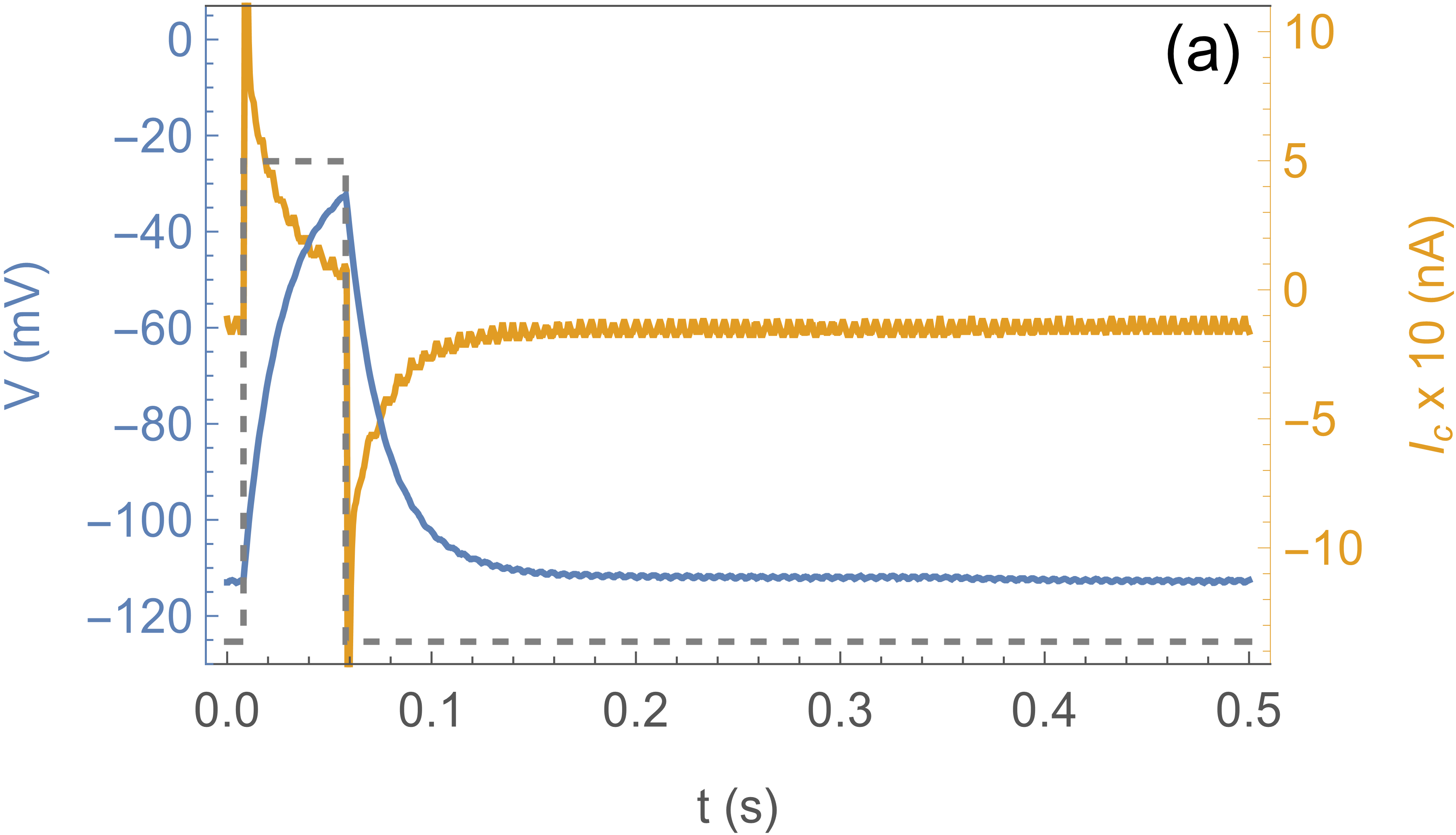}
\includegraphics[width=3in]{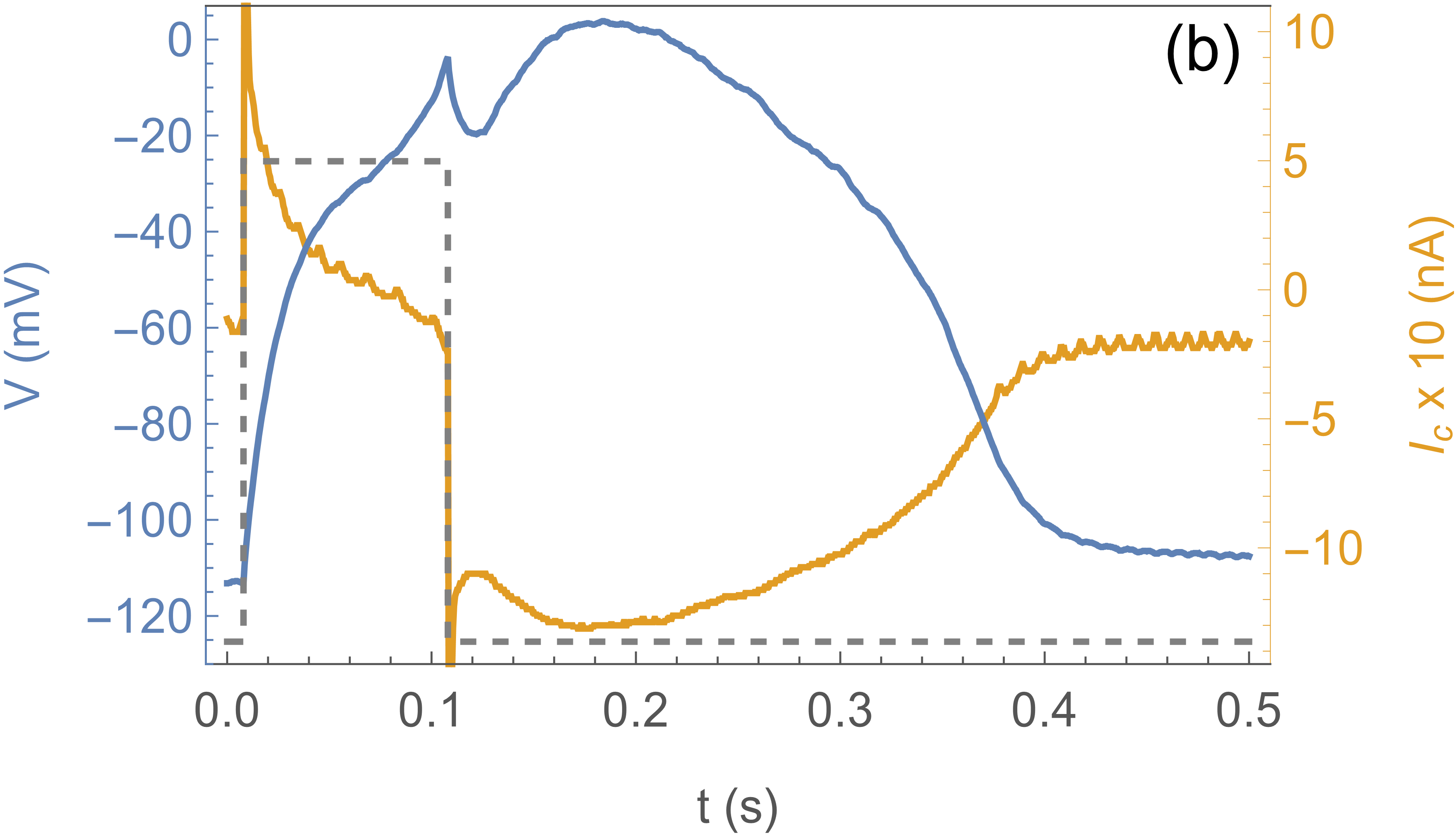}
\captionsetup{justification=raggedright, singlelinecheck=false}
\caption{Response of the artificial axon to a sub threshold (a) and above threshold (b) stimulus. The dotted line represents the input (the CLVC voltage protocol), the blue trace is the measured axon voltage $V (t)$, and the yellow trace is the current through the CLVC. In both (a) and (b) the CLVC protocol is a square pulse from $- 125 \, mV$ 
to $- 25 \, mV$, but the pulse width is $50 \, ms$ in (a) and $100 \, ms$ in (b), the latter being sufficient to 
induce firing. The CLVC resistance is 
$R_c = 100 \, M\Omega$, and from the sub threshold $V(t)$ trace, fitting with eq. (2) with closed channels 
($p_O = 0$), we find the values $C = 192 \, pF$ for the membrane capacitance and $R_{\ell} = 1.2 \, G\Omega$ 
for the leak resistance. }
\label{fig:threshold}
\end{figure}
\noindent We now proceed to build more features into the system. In electrophysiology, action potentials 
are usually evoked using a current clamp to inject the stimulus \cite{Hille}. This is an electronic analogue 
for the synaptic input. Fig. 4 shows a spike train similarly obtained stimulating the artificial 
axon with a current clamp, injecting a constant current $I_c \approx 100 \, pA$. However, due to the inactivation 
properties of the KvAP, to obtain this dynamics 
we must add a ``trigger'' function to the CLVC, which has the role of the missing second ionic species 
and channels. Namely, when the axon voltage $V(t)$ exceeds a fixed trigger voltage $V_T$ the CLVC switches from its fixed potential $V_{C1}$ to a more negative potential $V_{C2}$, for a fixed time $\tau$, then goes back to $V_{C1}$. 
This is necessary in order that the KvAP channels recover from inactivation. \\
The firing rate is a function of the 
input current: Fig. 5 shows the firing rate measured on spike trains as in Fig. 4, for increasing input current. 
All other parameters ($V_T$, $V_{C1}$, $V_{C2}$ and $\tau$) are fixed. The firing rate first increases linearly 
with input current, but will eventually saturate due to the constant width of the falling edge of the ``action potential''. This behavior is analogous to firing rate observations in patch clamp experiments of real neurons with constant current clamp inputs.  \\
\begin{figure}
\centering
\includegraphics[width=3in]{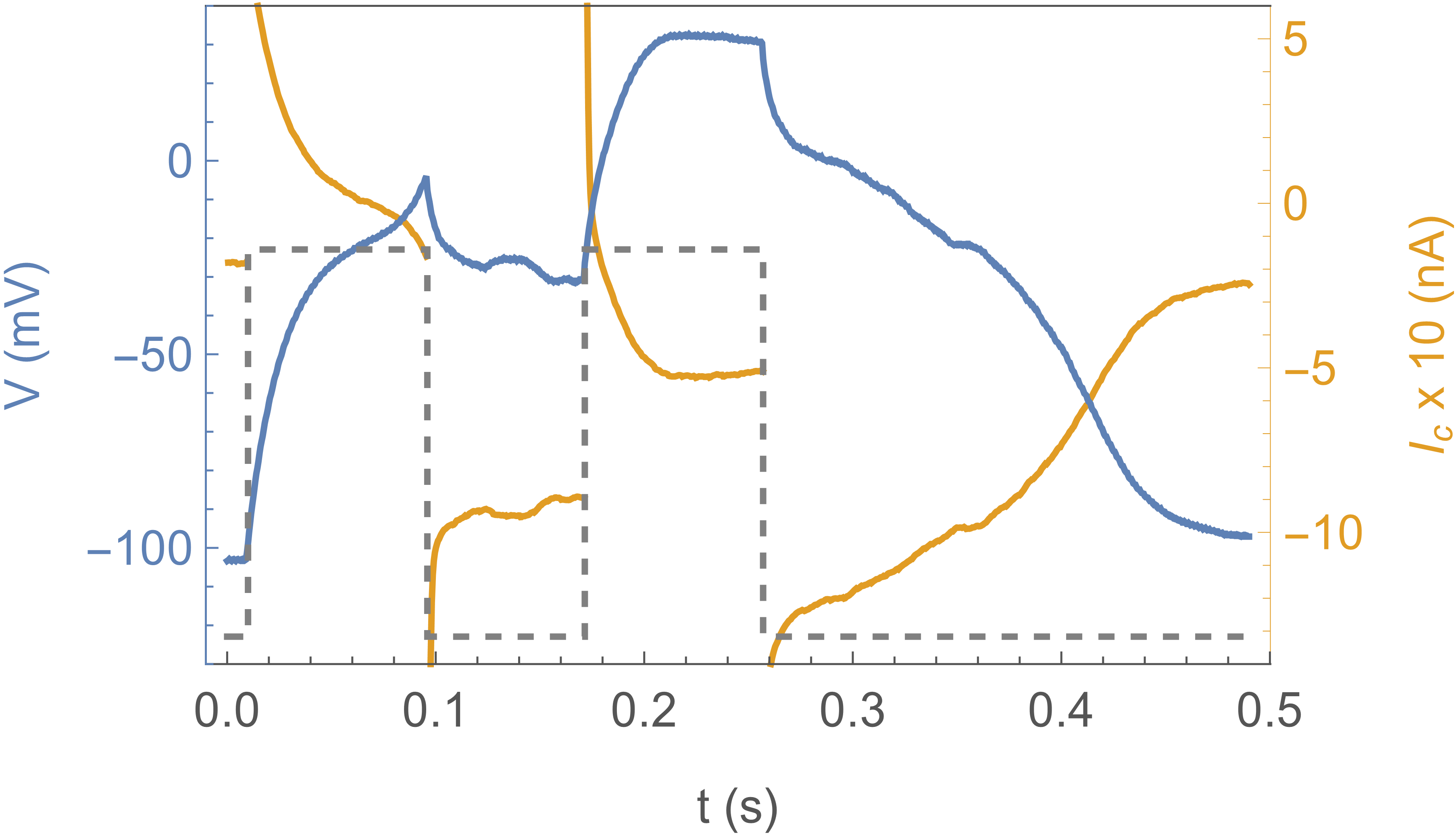}
\captionsetup{justification=raggedright, singlelinecheck=false}
\caption{Coincidence detection with the artificial axon. Two sub threshold pulses (dotted line) in close 
proximity cause the axon to fire. The first pulse induces opening of some channels, but the channel current 
cannot overwhelm the clamp (which has the role of the $Na^+$ gradient in the real neuron) absent the 
second pulse. The CLVC protocol is two identical pulses from $- 123 \, mV$ to $- 23 \, mV$, of width 
$86 \, ms$ and $80 \, ms$ apart. The CLVC resistance is $R_c = 100\,M\Omega$, and fitting to the model 
(2) (see text) gives the values $C = 210 \, pF$ and $R_{\ell} = 0.7 \, G \Omega$. }
\label{fig:coincidence}
\end{figure}
Next we connect two artificial axons through an electronic ``synapse'' which is a current clamp 
injecting into axon 2 a current controlled by the voltage $V_1 (t)$ in axon 1. It is a simple analogue 
circuit consisting of 220 op-amps and 640 resistors. In Fig. 6 we show action potential propagation from axon 1 
(blue trace), through the synapse, to axon 2 (yellow trace). Axon 1 is excited by a current clamp injecting a constant current 
$I = 90 \, pA$ for $0.1 < t < 3.3 \, s$. The hardwired synapse takes as input the voltage of axon 1, $V_1 (t)$, 
and injects into axon 2 a current $I_s$ proportional to the input $V_1 (t)$ when $V_1 (t) > 0$ : 
$I_s (t) = \alpha \, V_1 (t) \, \theta (V_1)$ with $\alpha \approx 2 \, pA / mV$ and $\theta$ the step function. 
There are no triggers in this experiment. 
We see that as $V_1 (t)$ crosses zero (at $t \approx 0.6 \, s$), $V_2 (t)$ starts to rise as the synapse injects current. 
Axon 2 ``fires'' (channels open) at $t \approx 1.3 \, s$. At $t \approx 1.5 \, s$ $V_1 (t)$ crosses zero on its way 
down; even though the synapse stops injecting current, $V_2 (t)$ remains high because the channel current 
overwhelms the CLVC current. Eventually channels inactivate and the CLVC pulls the membrane potential 
back down to the resting potential. In short, Fig. 6 demonstrates signal propagation between two axons connected 
by a synapse: the basic unit for a network. \\
{\bf Model.} The dynamics of the artificial axon is captured by the classic  
HH model of action potentials (as well as by simplified versions, discussed in a forthcoming publication). 
There are three contributions to the current in the system: the ion channel 
current $I_{Kv} = - N_0 \chi p_O (V - V_N)$ where $N_0$ is the number of channels, $\chi$ the open 
channel conductance, $p_O$ the probability that channels are open, $V_N$ the Nernst potential, and 
$V(t)$ the actual axon potential; the leak current $I_{\ell} = - N_0 \chi_{\ell} (V - V_N)$, $\chi_{\ell}$ 
being the leak conductance; the clamp current $I_{CLVC} = - (V - V_{CLVC}) / R_C$ where $R_c$ is 
the resistance limiting the current in the CLVC. These currents charge the membrane capacitance $C$: 
\begin{equation}
\frac{d V}{d t} \, = \, - \frac{N_0 \chi}{C} (p_O + \frac{\chi_{\ell}}{\chi}) [V(t) - V_N] \, - \, 
\frac{1}{R_C C} [V(t) - V_{CLVC}]
\label{eq:model1}
\end{equation}
If channel opening and closing was much faster than all other time scales in the system, $p_O$ would 
simply be the equilibrium open probability at the present voltage: $p_O = p(V)$ \cite{Amila2016}. 
However, this is not the case and one has to consider channel dynamics, which in HH is represented 
by a set of rate equations. The standard model has 3 states with respect to conduction: 
Closed, Open (probability $p_O (t)$), Inactive (probability $p_I (t)$), and several ``internal'' states. Namely, 
the ion channel being a tetramer, each subunit is assigned two states, 
$c_\alpha$ and $c_\beta$. The closed state that is accessible to the open and inactive states requires all 4 subunits be in the $c_\alpha$ state, associated with probability $p_4 (t)$. Probabilities $p_3 (t)$, $p_2 (t)$, $p_1 (t)$ and $p_0 (t)$ are associated with states inaccessible to the open and inactive states. These probabilities correspond to 3, 2, 1, and 0 subunits in the $c_\alpha$ state, respectively (more details are given in Supplementary Materials). The dynamics of $p_O$ is then given by the rate equations \cite{Hille,SingleChannel}: 
\begin{equation}
\begin{cases}
p'_O (t) \, = \, p_4 (t) k_{CO} (V) - p_O (t) k_{OC} (V) \\
p'_I (t) \, = \, p_4 (t) k_{CI} (V) - p_I (t) k_{IC} (V) \\
p'_4 (t) \, = \, p_O (t) k_{OC} (V) - p_4 (t) [k_{CO} (V) + 4 \beta (V) + k_{CI} (V)]  \\
\indent\indent\indent\indent\indent\indent\indent\indent\indent\indent\indent + p_I (t) k_{IC} (V) + p_3 (t) \alpha (V) \\
p'_3 (t) \, = \, 4 p_4 (t) \beta (V) - p_3 (t) [1 \alpha (V) + 3 \beta (V)]  \\
\indent\indent\indent\indent\indent\indent\indent\indent\indent\indent\indent\indent\indent\indent\indent\indent\indent + 2 p_2 (t) \alpha (V)  \\
p'_2 (t) \, = \, 3 p_3 (t) \beta (V) - p_2 (t) [2 \alpha (V) + 2 \beta (V)]  \\
\indent\indent\indent\indent\indent\indent\indent\indent\indent\indent\indent\indent\indent\indent\indent\indent\indent + 3 p_1 (t) \alpha (V) \\
p'_1 (t) \, = \, 2 p_2 (t) \beta (V) - p_1 (t) [3 \alpha (V) + 1 \beta (V)] \\
\indent\indent\indent\indent\indent\indent\indent\indent\indent\indent\indent\indent\indent\indent\indent\indent\indent + 4 p_0 (t) \alpha (V) \\
p'_0 (t) \, = \, 1 p_1 (t) \beta (V) - 4 p_0 (t) \alpha (V) \\
\end{cases}
\label{eq:model2}
\end{equation}
and eqs. (\ref{eq:model1}) and (\ref{eq:model2}) constitute the model. For the voltage dependence of the rates in (\ref{eq:model2}) one takes exponential (Arrhenius) functions; the corresponding parameters have indeed been measured for KvAP \cite{Schmidt2009}. We use parameters that best fit our system. 
In Fig. 6, the fit superimposed on the measured voltage trace $V_1 (t)$ shows an example 
of reproducing the dynamics seen in the experiments using this model. There are two interesting aspects. One is that the model can be used as an engineering tool to predict how more complicated 
configurations of artificial axons will behave. The other is the nonlinear dynamics of this 
interesting complex system, which however, has been studied \cite{Koch}.  \\
\begin{figure}
\centering
\includegraphics[width=3in]{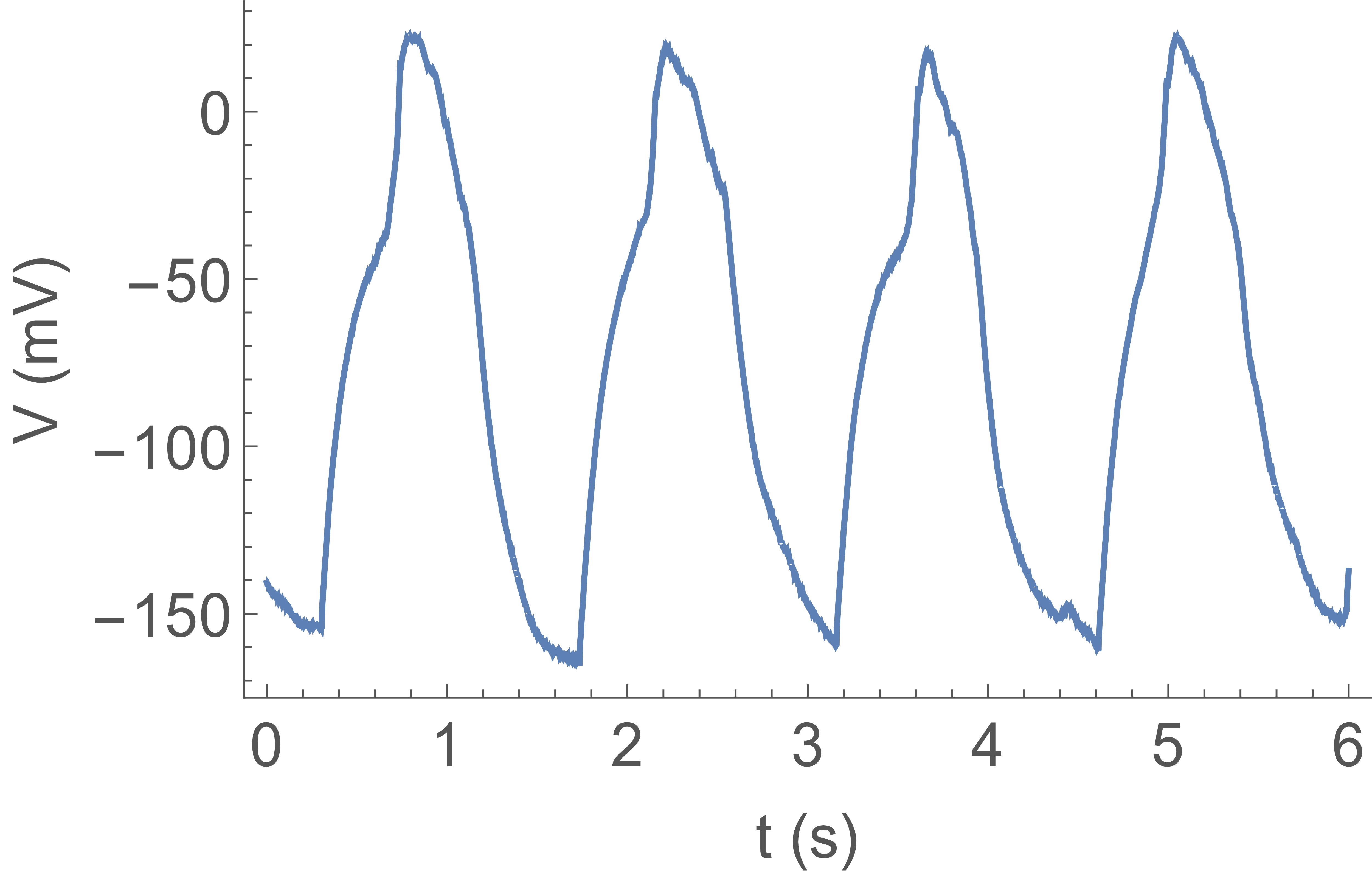}
\captionsetup{justification=raggedright, singlelinecheck=false}
\caption{Spike train obtained under constant input current conditions. A current clamp injects 
$I_c = 100 \, pA$ into the axon. The CLVC parameters are $V_T = 0 \, mV$, $V_{C1} = - 364 \, mV$, 
$V_{C2} = - 636 \, mV$, $\tau = 1 \, s$ (see text). This particular prep has a relatively large leak current. }
\label{fig:multifiring}
\end{figure}
\begin{figure}
	\centering
	\includegraphics[width=3in]{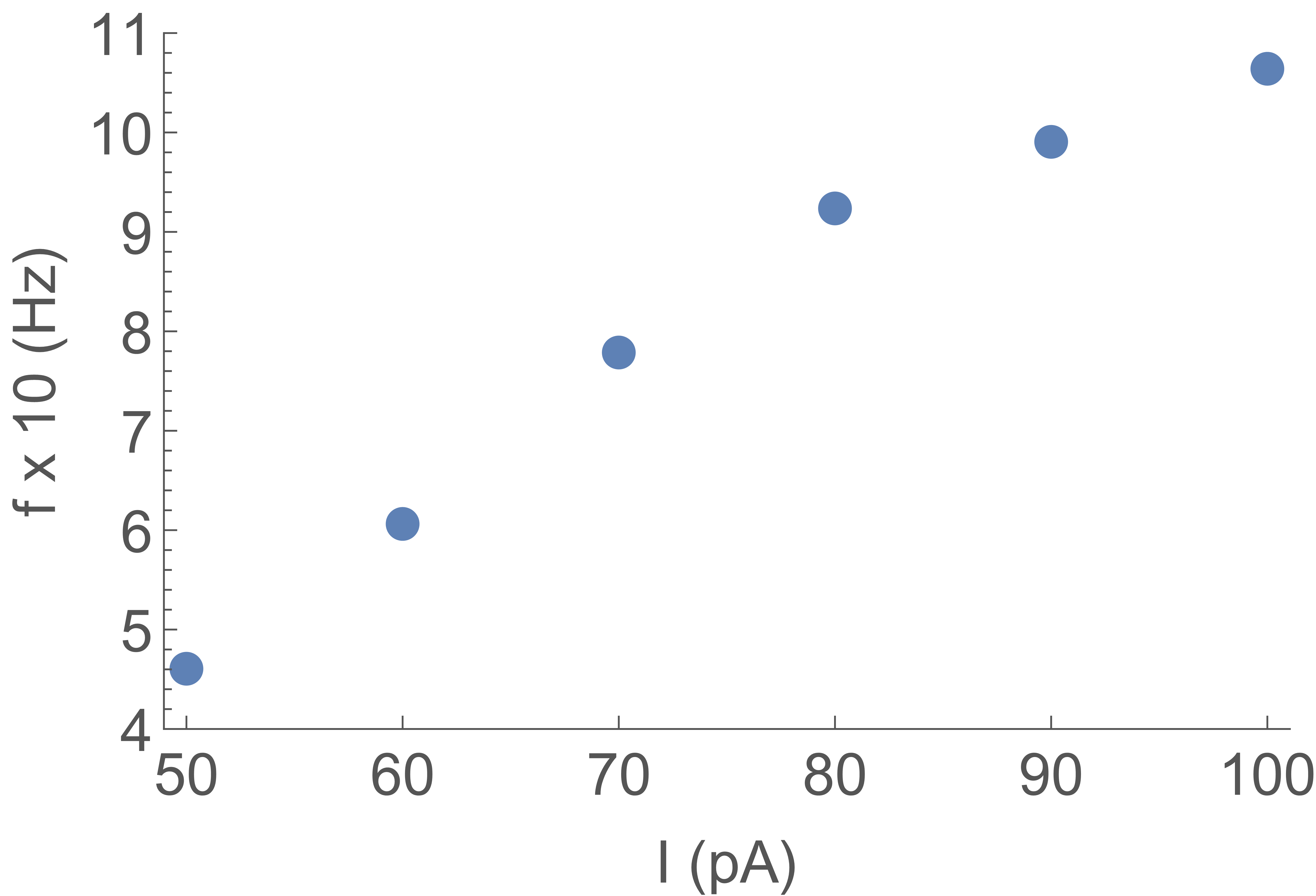}
	\captionsetup{justification=raggedright, singlelinecheck=false}
	\caption{Measured firing rate vs input current. Each data point is obtained from a spike train as in Fig. 4. 
The fixed CLVC parameters are:  $V_T = 14 \, mV$, $V_{C1} = - 125 \, mV$, $V_{C2} = - 459 \, mV$, 
$\tau = 500 \, ms$. }
	\label{fig:rates}
\end{figure}
{\bf Discussion.} The artificial axon represents an experimentally controlled, synthetic realization of this most 
interesting of nonlinear dynamical systems, the excitable cell. 
It is to our knowledge the first artificial system based on the bio-components that generates 
something similar to action potentials. Even in its present simple form it can 
serve as a breadboard for electrophysiology measurements which does not involve lab animals. For example, 
the electrophysiology of real neurons is extremely temperature dependent \cite{Katz1949, Prince1985}; future experiments with the artificial axon may elucidate the causes. 
\\ 
Networks of excitable nodes are intensively studied computationally \cite{Shew2014}. What is the outlook for a network of artificial axons? On the positive side there are power and space 
considerations. Reasoning on a hypothetical $A = 100 \, \mu m^2$ membrane patch with $N = 100$ 
channels, responding on a time scale $\Delta t \sim 10 \, ms$ (our present axon is larger: 
$A \sim 10^4 \, \mu m^2$, and slower: $\Delta t \sim 100 \, ms$, but size can be reduced and much 
faster channels than the KvAP are available, so the above are by no means unrealistic conditions), the 
capacitance is $C \approx 10^{- 13} \, F$ and it takes only the transfer of $\sim 10^5$ charges to step 
the voltage by $\sim 100 \, mV$, giving a dissipation of $\sim 10^4 \, eV$ over $10 \, ms$, or 
$0.1 \, pW$. One can certainly imagine a $\sim 10 \, cm$ size chip containing $\sim 10^8$ membrane 
patches. For comparison, our brain has $\sim 10^{11}$ neurons and consumes $\sim 30 \, W$, or 
$300 \, pW$ per neuron, though only a fraction of this dissipation is due to action potentials. 
A flip-flop circuit in a CPU (consisting of $\sim 10$ transistors) consumes $\sim 100 \, \mu W$ at 
$1 \, GHz$ or $\sim 10 \, pW$ at $100 \, Hz$. This factor $100$ in power dissipated  
compared to our hypothetical membrane patch 
comes essentially from the factor $10$ reduction in operating voltage: $1 \, V \rightarrow 100 \, mV$ 
between transistors and voltage gated channels. The biological channels {\it are} the much sought for 
low voltage transistors! Indeed, this mismatch is the main problem while interfacing ``bare'' electronic 
components with the artificial axon; for instance, in order to make a one-way synapse one cannot 
simply connect two axons through a diode, because of the voltage drop across the diode. \\
Now on the negative side, for a viable network there are problems to be solved requiring altogether 
new inventions. The first is scaling up the system, introducing two channel species, and obtaining the 
necessary robustness. Another challenge is creating an artificial electrochemical synapse which uses 
only low voltage components. In the next decades, we look forward to progress being made on both fronts. \\
\begin{figure}
	\centering
	\includegraphics[width=3in]{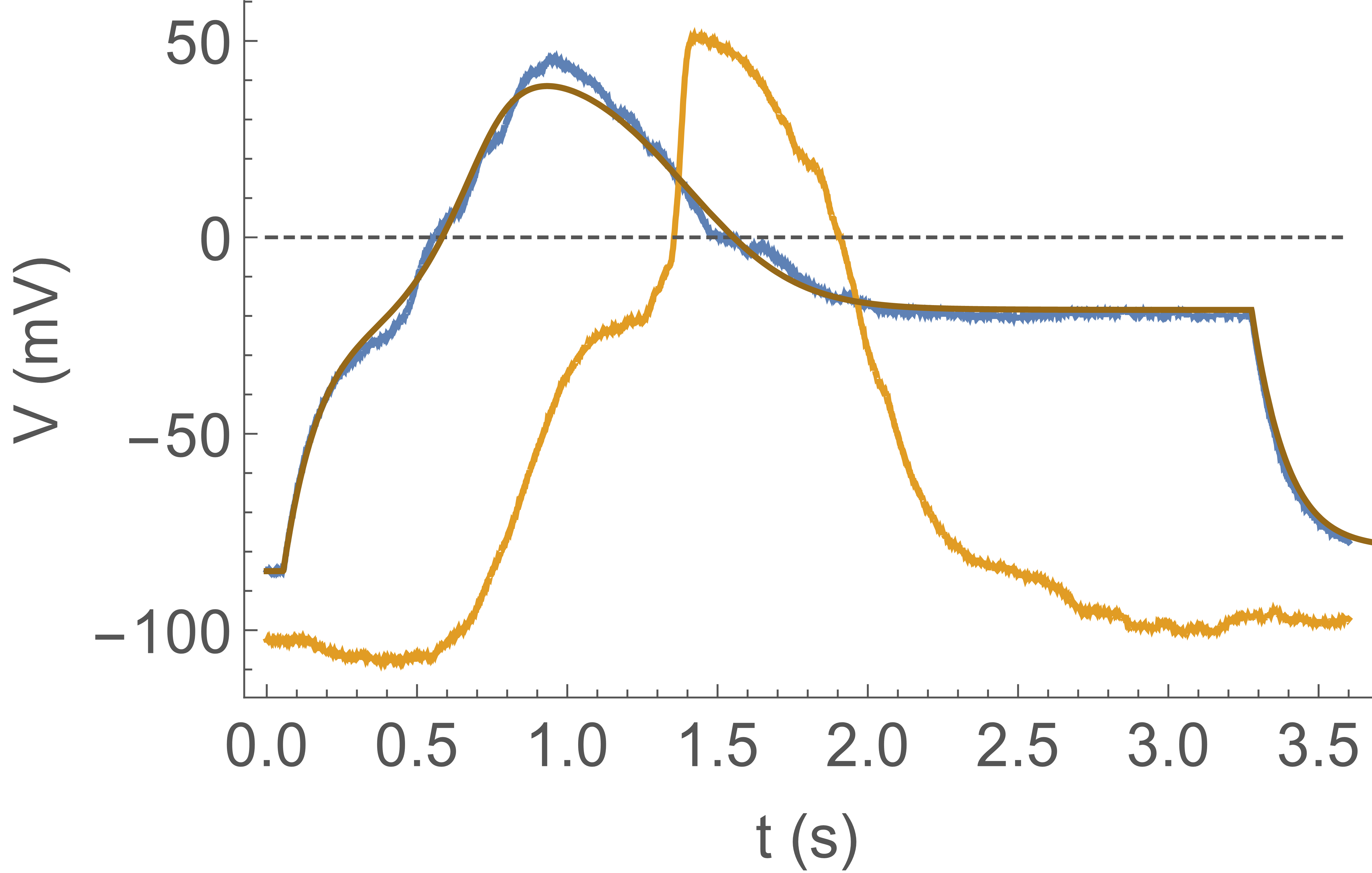}
	\captionsetup{justification=raggedright, singlelinecheck=false}
	\caption{Signal propagation from axon 1, through the synapse, to axon 2. The synapse is a hard wired 
current clamp controlled by $V_1 (t)$ and injecting current into axon 2. Also shown is a fit of the axon 1 
potential $V_1 (t)$ using the model (2), (3). It returns the parameter values $N_0 = 461$, 
$C = 175 \, pF$, $\chi = (6G\Omega)^{-1}$, $\chi_{\ell} = (68G\Omega)^{-1} / N_0$, $V_N = 56 \, mV$. 
The CLVC resistance was $R_c = 2\,G\Omega$.}
	\label{fig:synapse}
\end{figure}

\section{Acknowledgements}
\noindent We thank Roderick MacKinnon for the original gift of the KvAP plasmid, 
and Amila Ariyaratne for invaluable advice and support.  
This work was supported by NSF grant DMR-1404400.

\bibliography{Hector_1}

\onecolumngrid
\pagebreak
\begin{center}
	\textbf{\large Supplemental Materials}
\end{center}

\twocolumngrid
\maketitle

\section{Materials and Methods}

We use a voltage gated potassium ion channel from the thermophilic bacterium {\it Aeropyrum pernix} (KvAP). The channel is closed at large negative voltages; the midpoint of the open probability curve is at $V\approx-20\,mV$, and the open channel conductance is $\chi \approx 10\,pA/60\,mV$ [8, 9]. KvAP has a slow dynamics of inactivation; to recover from inactivation, it must be held at voltages below $\sim-100\,mV$. \\
Vesicles and decane lipid membranes are made with the phospholipid DPhPC.

\subsection{A. Expression and Reconstitution}

KvAP protein expression, purification and reconstitution mostly follow the protocols in [9], and are described at length in [8]. Briefly, the KvAP gene in vector pQE60 is expressed in {\it E. coli} cells (Aligent XL-1 Blue). Vector transformation into {\it E. coli} is achieved with a 45 second, 42 degree heat pulse. Cells with KvAP protein are re-suspended in buffer and lysed in a homogenizer (Avestin EmulsiFlex-C3). Protein is extracted from the lysate with decylmaltoside detergent (DM, Anatrace) and centrifugation. Talon nickel bead resin (Clonetech) is added to the his-tagged ion channel solution for binding and purification. The mixture is passed through a Talon affinity column then eluted. The elution is treated with thrombin to remove the his-tag from the ion channel. Thrombin and his-tag are removed from the sample by size exclusion chromatography (GE Healthcare, Superdex-200).\\
\noindent DPhPC is suspended in buffer and sonicated to produce uni-lamellar vesicles. The protein sample is added to the vesicle solution then passed through 3 spin desalting columns (Thermo Scientific, Zeba) to remove most DM from solution. DM in complexes with lipid are removed with detergent absorbing bio beads (Bio-Rad). After DM removal, the vesicle-channel solution is aliquoted and flash frozen in an ethanol-dry ice bath, then stored at -80 C.

\subsection{B. Electrophysiology Setup}

\noindent Overall setup, including headstage and voltage amplifiers are described at length in publications [3] and [8], and data acquisition is described in publication [3]. Setup and components will be described briefly.\\
\noindent A $ \sim100 \mu m$ aperture on a plastic cup is used as the support for a decane lipid membrane. The membrane separates two chambers of different $K^+$ concentrations. One chamber is grounded, through an Ag/AgCl electrode. The other contains two Ag/AgCl electrodes (see Fig. 1), one for the current limited voltage clamp (CLVC), the other connecting to:  1) an amplifier for the measurement of the axon voltage $V(t)$; 2) an independent current clamp used to inject a stimulus current; 3) the input of the electronic synapse feeding into axon 2 (for the two-axons experiment). For the CLVC, we use the resistor values $R_c = 100M\Omega$ and $2G\Omega$ for different situations. For the two-axon measurements, both wells are connected to independent CVLC’s. All electronics are made in-house.\\
\noindent The electronic synapse connecting two axons is a current clamp. The membrane voltage $V_1 (t)$ of axon 1 is the command voltage for the synapse, effectively assigning axon 1 the role of the pre-synaptic neuron. The synapse can inject current into well 2 only when $V_1 (t) >0$. This switch behavior is achieved with a solid state relay [3] that is switched to open/close with an op-amp. Positive/negative $V_1 (t)$ controls the polarity of op-amp saturation, which in turn switches the relay to the open/close position. Design figures and detailed descriptions are to be provided in later publications.\\

\subsection{C. Modeling}

\noindent The scheme for KvAP gating used to fit the channel behavior is from [6]. We used as a guideline the rate constants in [6] for DPhPc lipid membranes, with modifications to fit our data. All rate constants have the form $k = k_0 e^{-zV}$. The gating scheme and $k_0$, z values are provided in Supplemental Material Figure 1.

\begin{figure}
	\centering
	\includegraphics[width=3in]{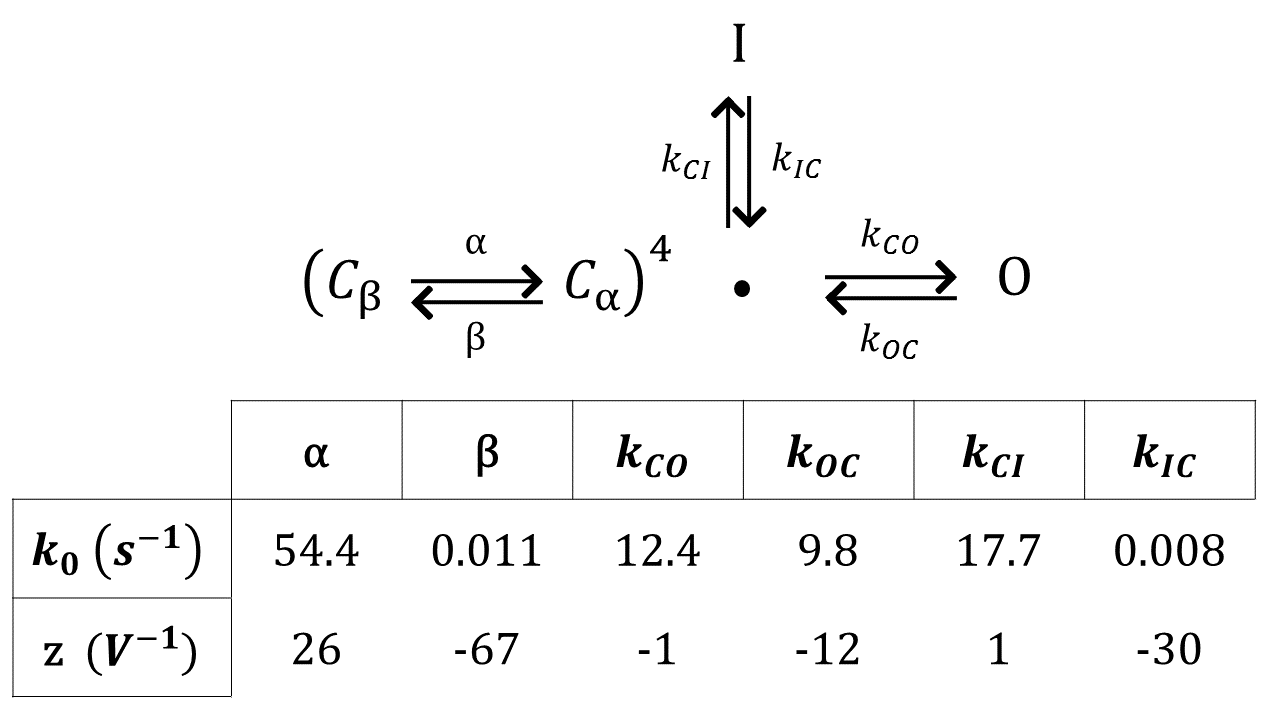}
	\captionsetup{justification=raggedright, singlelinecheck=false}
	\caption{Gating scheme and rate constant fit values for channel behavior.}
	\label{fig:coincidence}
\end{figure}

\end{document}